# Characteristics of SEPs during Solar Cycle 21-24

Raj Kumar*, Ramesh Chandra, Bimal Pande, Seema Pande

Department of Physics, DSB Campus, Kumaun University, Nainital-263001
*E-mail: rajkchanyal@gmail.com

**Abstract:** The study of the solar energetic particle events (SEPs) and their association with solar flares and other activities are very crucial to understand the space weather. Keeping this in view, in this paper, we present the study of the SEPs (intensity ≥ 10 pfu) during the solar cycle 21 to 24 (1976-2017) in > 10 MeV energy channels associated with solar flares. For our analysis, we have used the data from different instruments onboard SOHO satellite. We have examined the flare size, source location, CMEs characteristics of associated SEPs. About 31% and 69% of the SEPs were originated from the eastern and western solar hemisphere respectively. The average CME speed and width were 1238 $kms^{-1}$ and $253^0$ respectively. About 58% SEPs were associated with halo CMEs and 42% of SEPs associated with CMEs width varying from $10^0$ to $250^0$ respectively.

## 1. Introduction

Solar eruptions release high energetic particles, which possess the energy ranging from few tens of keV to thousands of GeV. These high energetic particles are called solar energetic particles (SEPs). These high energetic particles can reach our space–weather depending upon their location on solar surface and their direction of eruption (for detail see the review by Desai and Giacalone, 2016). If the energy of particles reaches the GeV energy range, then it is termed as ground level enhancements (GLEs).

There are two possible mechanisms responsible for the generation/origin of SEPs namely shock wave driven and magnetic reconnection. Let us discuss these mechanisms briefly as follows:

According to Reames (1999), SEPs can be generated by the shock waves driven by coronal mass ejections (CMEs). CME can generate the shock waves when the speed of CMEs is greater than the coronal sound speed. Later on many authors confirmed it (Giacalone 2012, Reames 2012, Gopalswamy *et al.* 2015). Strong SEPs show close association with halo CMEs and the average speed of the associated CMEs are ≈ 1500 $kms^{-1}$ (Gopalswamy *et al.* 2014). Stronger shocks are driven by the high speed CMEs. Therefore, correlation between CME speed and SEP intensity should be strong. However, Gopalswamy *et al.*, (2004) found in many cases that with the occurrence of SEP event there was no corresponding type II radio burst. Solar type -II radio bursts are due to shocks generated ahead of the CMEs during eruptions. In addition to this, the correlation between CME speed and SEPs intensity is not found strong. The possible reason for this weak association was discussed in the work of Gopalswamy *et al.* 2012. According to their study this poor correlation could be due to CME-CME interaction, deflection of CMEs from the coronal hole regions etc.

Another mechanism responsible for the generation of SEPs is solar flares which are due to magnetic reconnection (Mason *et al.*, 1999,

Miteva *et al.* 2013). SEPs generated by solar flares are impulsive and short lived in nature (Cane *et al.*, 1986; Reames, 1999; Laurenza *et al.*, 2009; Cane *et al.*, 2006; Bhatt *et al.*, 2013, Chandra *et al.*, 2013). Association of the SEPs with reconnection can be confirmed by their association with type III radio bursts. If SEP events are associated with type III radio bursts it means charged particles can escape to the interplanetary space (Wild, Smerd and Weiss, 1963, Buttighoffer, 1998, Cane, Richardson and von Rosenvinge, 2010, Miteva *et al.*, 2014, Winter and Ledbetter, 2015, references cited therein). Authors have studied in past the SEP events with the solar X-ray flares and found a weak correlation between SEP intensity and GOES X-ray flux (Gopalswamy *et al.* 2003, Trottet *et al.* 2015).

It is believed that both the shock and flare occur during the solar eruptions. Therefore it is difficult to conclude which mechanism is responsible for the generation of SEPs.

In addition to above two discussed mechanisms, there are studies when the SEPs production depends on the properties of active region (Thalmann *et al.*, 2015, Bronarska & Michalek, 2017, Mitra *et al.*, 2018, Bruno *et al.*, 2019, Kashapova *et al.*, 2019, ). These cited authors found the active regions, which are more magnetically complex produce the strong SEPs. Recently Kouloumvakos *et al.* (2015) did a statistical study of SEPs and radio type II and type III bursts. They found majority of cases were associated with both type II and III radio bursts. However, the type III radio bursts association was stronger.

From the above discussion, it is clear that the SEP production is still the subject of debate. Therefore, in this paper, we present a comparative study of SEP events (intensity ≥ 10 pfu) in > 10 MeV energy channel with the different properties of CME and GOES X-ray flares during solar cycle 21-24 (1976-2017). The paper is organized as follows: section 2 contains the description of data sets. The analysis and results are described in section 3. Section 4 presents the summary of the paper.

## 2. Data Sets

For the present study, we have taken the SEP events in the ≥ 10 MeV energy channel during solar cycle 21-24 (1976-2017) associated with solar flares. The proton intensity is in particle flux unit (pfu) (1 pfu is equal to 1 proton $cm^{-1}$ $s^{-1}$ $sr^{-1}$). The total number of SEP events during this period is 270. We have also counted the no. of SEPs during solar cycle 21, 22, 23 and 24. The total SEPs during 21, 22, 23 and 24 are 63, 73, 91 and 43 respectively. We have used all the observed GOES X-ray flare classes corresponding to the associated SEP events. The GOES data are downloaded from http://sec.noaa.gov.

We have taken the CME data from the Large Angle and Spectrometric Coronagraph (LASCO, Brueckner *et al.*, 1995) onboard Solar and Hemispheric Observatory (SOHO) satellite as cataloged at the Coordinated Data Analysis Workshops (CDAW) data center (http://cdaw.gsfc.nasa.gov, Gopalswamy *et al.*, 2009). The CMEs source location is taken using images and movies available at SOHO/LASCO CME catalogue (soho/lasco cme catalogue/java movies/). Since the CME observations are available from 1997 onwards, hence the analysis before 1996 does not include the CME data. A complete table of data set used is given in the appendix.

An example of a SEP during solar cycle 24 is displayed in Figure 1. The source of this SEP is located at N14E02 on solar disc. The event was associated with a halo CME having speed ≈ 1425 $kms^{-1}$. This event is associated with GOES X1.6 class flare.

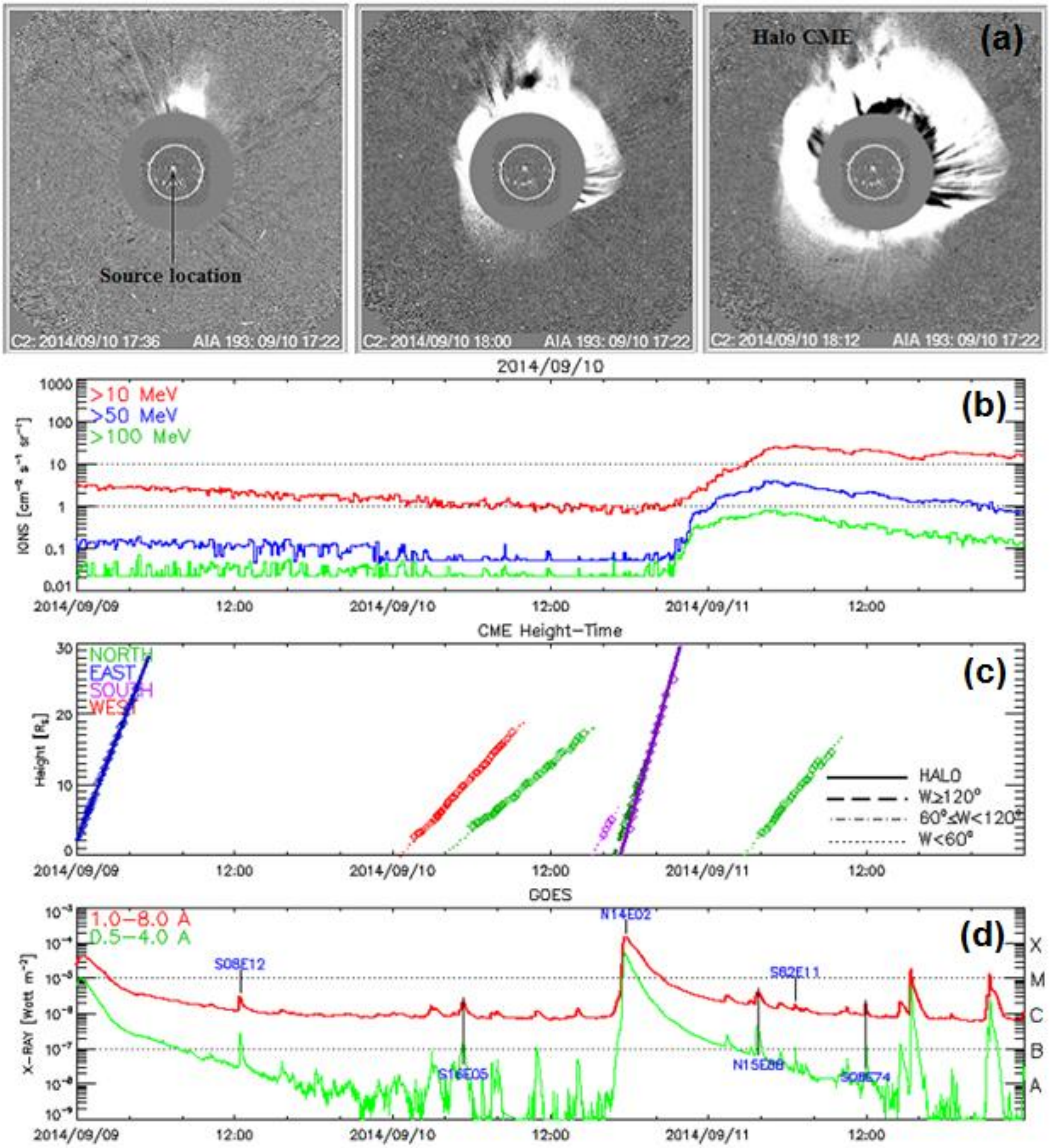

Figure 1. An example of SEP event associated with halo CME and its source region on September 10, 2014. Panel 'a' shows the three snap shots of halo CME that produced the SEP event. Panel 'b' is showing the time variation of SEP intensity in three energy channels. Panels '**c**' and'd' are showing the CME Height-Time plot and GOES soft X-ray flux in two energy channels, respectively (taken from SOHO/LASCO catalog).

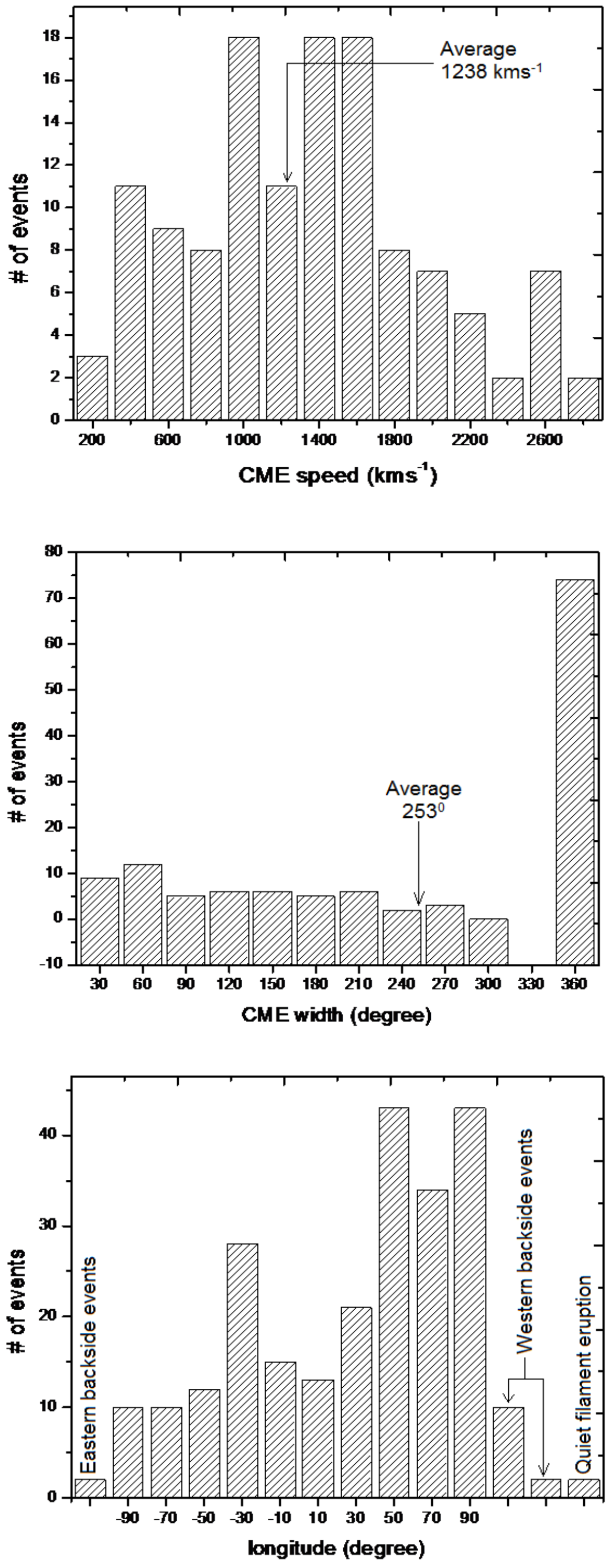

Figure 2. Histogram of number of SEP events in view of CME characteristics: Speed in $kms^{-1}$ (top), width (middle) and source longitude (bottom).

## 3. Analysis and Results

Our study presents the statistical analysis of SEP events associated with solar flares during 21-24. The different statistical results are presented in following sections.

### 3.1 SEP events and CME characteristics

The numbers of SEP events with CME speeds is shown in Figure 2 (top). SEP events of speeds ranging from 1000 to 1600 $kms^{-1}$ are maximum. It is about 12% of the total number of events. The average speed of the CMEs is ≈ 1238 $kms^{-1}$. About 16% of the SEPs are associated with slow CME (speed ≤500 $kms^{1}$). 13% events are associated with very high speed > 2000 $kms^{-1}$.

The variation of CMEs width and number of events is presented in the middle of the Figure 2. CME width given here is simply 2D width based on plane-of-sky observations, while the actual 3D width, in cases, could be significantly different. Jang *et al.*, 2016 did a statistical study of comparison between 2D and 3D parameters of halo CMEs. They found that the 3D widths are much smaller than the 2D widths. The average value of CME width is about $253^0$. The maximum number of SEP events (58%) is associated with halo CMEs. About 42% of the CMEs are distributed over the width ranging from $10^0$ to $250^0$.

Figure 2 (bottom) displays the location of the sources in the eastern and western solar hemisphere. About 52% events are associated with the western solar hemisphere (longitude 20-$90^0$). 23% events are originated from the central region (± $20^0$). The region below $20^0$ eastern solar hemisphere produced 20% events. 5% events are originated from western backside solar hemisphere and 2 events from eastern backside solar hemisphere. 2 events are corresponding to quiet filament eruption. This is an agreement with the **scenario** that due to good magnetic connectivity most of the SEPs are originated from the western solar hemisphere.

### 3.2. SEP events and flare characteristics

As we have discussed in the introduction section the SEPs can be generated by the solar flares also. Therefore, we analyse here the relation between the flare class and the SEP events.

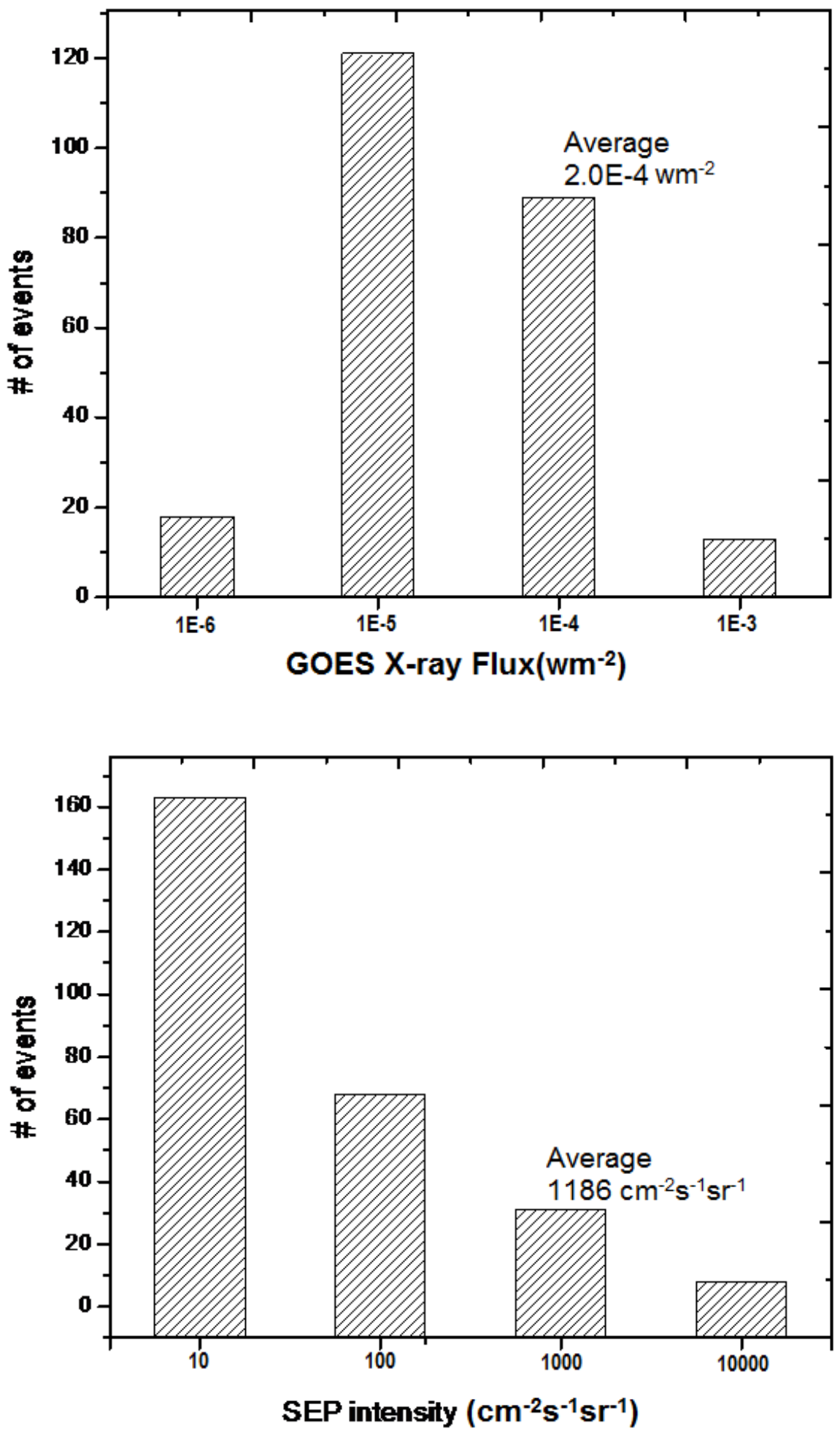

Figure 3. Histogram of number of SEP events in view of intensities: X-ray flux in $Wm^{-2}$ (top) and SEP intensity in pfu (bottom).

Figure 3 (top) depicts the number of SEP events and the GOES X-ray flux. We observed that the average flare intensity is $2.0x10^{-4}$ $Wm^{-2}$. We found 43% SEPs associated with X-class of solar flares. 49% and 8% SEPs are associated with M and C-class of solar flares, respectively.

SEP intensities are plotted against the number of SEP events in Figure 3 (bottom). The average SEP intensity is 1186 pfu. We have found about 60% of the SEP events have intensity 10 pfu. About 25% events have intensity between 10 to 100 pfu; 11% events in the range of 1000 pfu; events with intensity greater than 10000 pfu are 3%. It is inferred from this analysis that majority of SEPs are associated with M and X-class solar flares.

The number of C, M and X class flares is 18,121 and 102 respectively over the studies period. From solar cycle 21 to 24, number of GOES X – class flares varies as 30, 34, 27 and 11. For GOES M class, the numbers vary as 23, 32, 45 and 21 and the number of flares for GOES C class is 02, 03, 09 and 04. The occurrence rate of solar flares with solar cycles was studied in previous investigations (Howard 1974; Joshi & Pant 2005; Joshi *et al.*, 2006; Gao *et al.*, 2009; Joshi *et al.*, 2015 and references cited therein). Joshi *et al*., 2015 found that the number of C, M and X GOES class solar flares decreased with solar cycle. However, we did not found this decreasing trend in SEP events and associated solar flares. This inconsistency indicates that all solar flares are not associated with SEP events.

### 3.3. Correlation analysis

In this section we have analysed the correlation of SEP intensity with CME and GOES X-ray flux. The detailed description is given in the following subsections.

#### 3.3.1. CME speed and SEP intensity

Figure 4 presents the scatter plot between CMEs speed and SEP intensity. We compute the correlation coefficient between these two parameters and the value is 0.39.

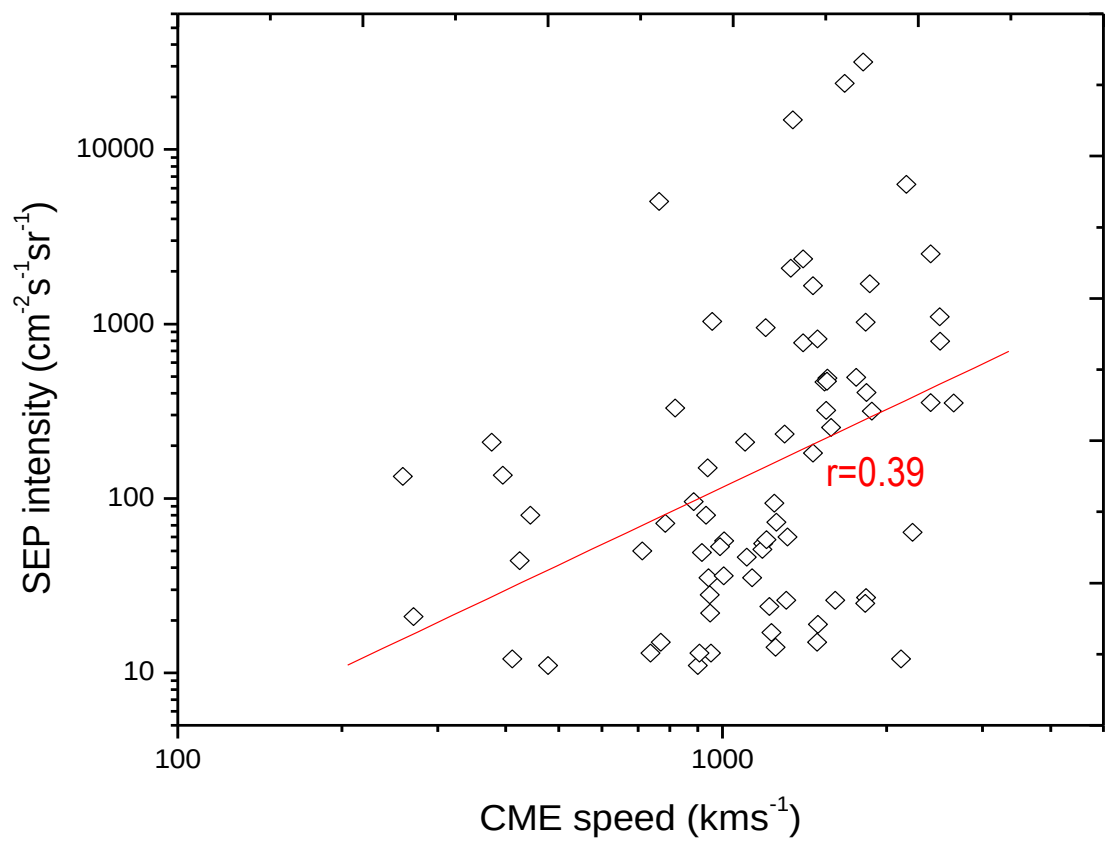

Figure 4. Scatter plot between CME speed and SEP intensity for all events.

This indicates the weaker correlation between these quantities. Further we have calculated the correlation coefficient for the longitude range i.e., eastern (20 to 70$^0$), central (± 20$^0$) and western (20 to 70 and 70 to 90$^0$). For our analysis, we have not considered the events originating from eastern hemisphere from longitude 70 to 90$^0$, because we found only four events within this longitude range. The results are presented in Figure 5. The correlation coefficient for longitude range from 20 to 70$^0$ in eastern hemisphere is 0.41. The correlation coefficient is 0.52 for the central region. As we move in the western hemisphere from the longitude 20 to 70$^0$, the correlation coefficient becomes poor with value 0.38. From longitude range 70 to western limb the correlation becomes maximum i.e., 0.53. This confirms the idea that when we move towards western limb, the magnetic connectivity becomes better as proposed in Parker spiraling. Our this result is in agreement with the previous findings (Kahler, 2001; Gopalswamy *et al.*, 2003, 2004 and references therein).

#### 3.3.2. Flare class and SEP intensity

The correlation between X-ray flare flux and SEPs intensity is presented in Figure 6. The top panel of figure presents plot for the events associated with all CMEs (including total CMEs width), while the bottom panel shows the plot for the events, which are associated with halo CMEs. The correlation coefficient for the total and the events associated with halo CMEs are 0.40 and 0.34 respectively. However, in both the cases, the correlation coefficient is poor, which support the earlier studies (Gopalswamy *et al.*, 2003, 2004, Chandra *et al*, 2013). Moreover, we observed that the correlation coefficient is decreased when we took only halo CMEs. This decrease in correlation may be explained by the fact that during the halo CMEs the contribution from the shock is dominated in comparison to the solar flares. Another possibility for this poor correlation could be the weak acceleration of particles during the flare time. Due to this weak acceleration the charged particles cannot reach the interplanetary space.

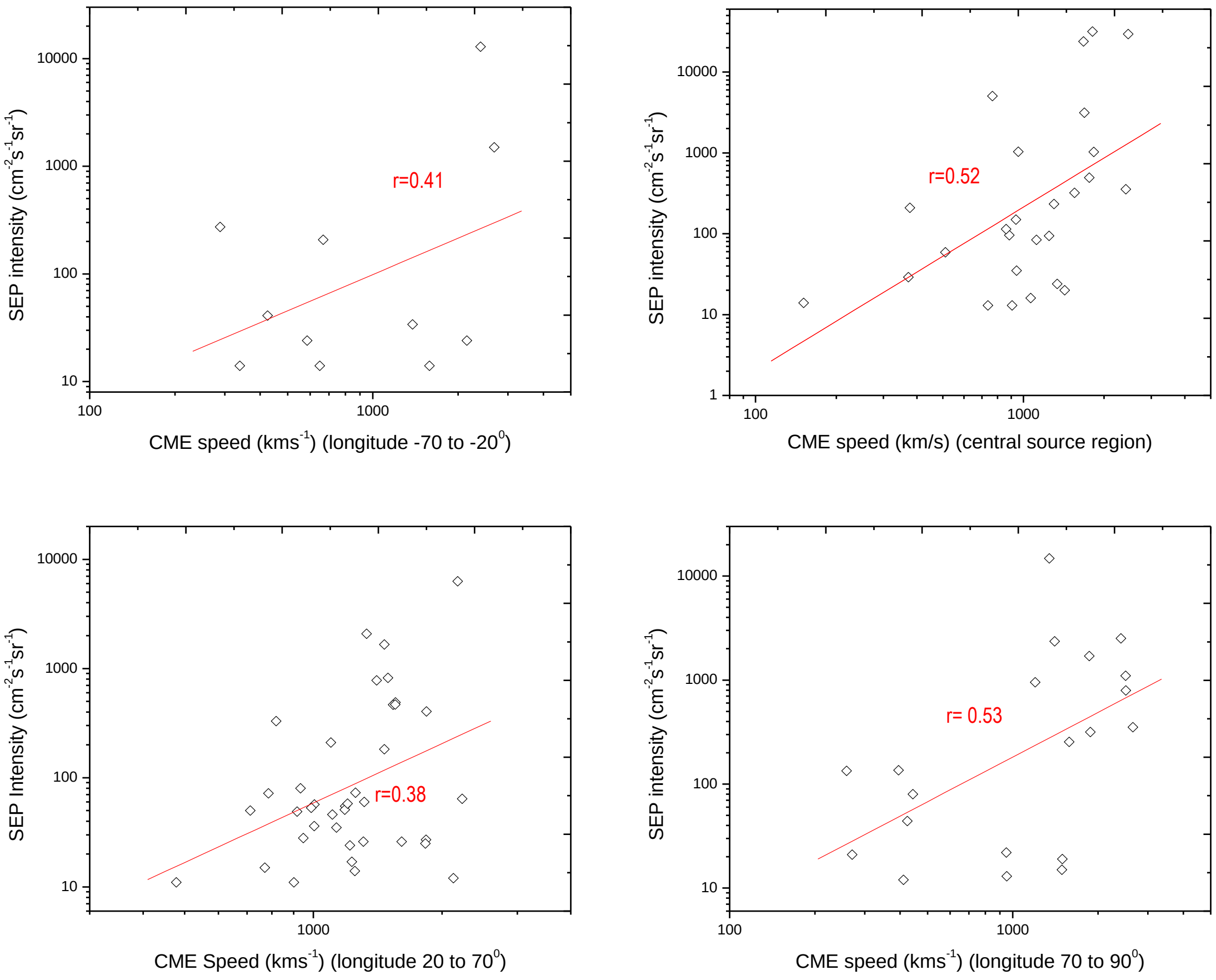

Figure 5. Scatter plots between SEP intensity and CME speed for different longitude zones of solar surface. Central zone is defined as the longitude $\pm 20^0$. The regression line and correlation coefficients (r) are shown in plots.

## 3.4 GLEs during Solar Cycle 23 and 24

We have selected some events from solar cycle 23 and 24, which produced ground level enhancements (GLEs) listed at http://gle.oulu.fi site. We chose this period of events because of the availability of LASCO CME data. This is given in Table 1. We examined the properties/characteristics of these GLE associated parameters. We found the majority of GLEs (77%) were associated with halo CMEs. The speed of all the associated CMEs was high ranging from 938 to 2598 kms$^{-1}$. For one event (August 24, 1998) the CME data was not available. 10 out of 13 associated flares were X-class and 02 were M-class flares. One event was not associated with any GOES X-ray flare. It was produced by filament eruption. However, the speed of associated CME was very high i.e. 2465 kms$^{-1}$. This speed indicates the production of strong shock during the CME propagation. All these GLE producing event sources were located either in central region or close to western limb. The properties of GLE events were discussed in detail by Gopalswamy *et al.* 2012. Richardson and Cane (2010) also made a study, which summarized different properties of the ICME during the solar cycle 23.

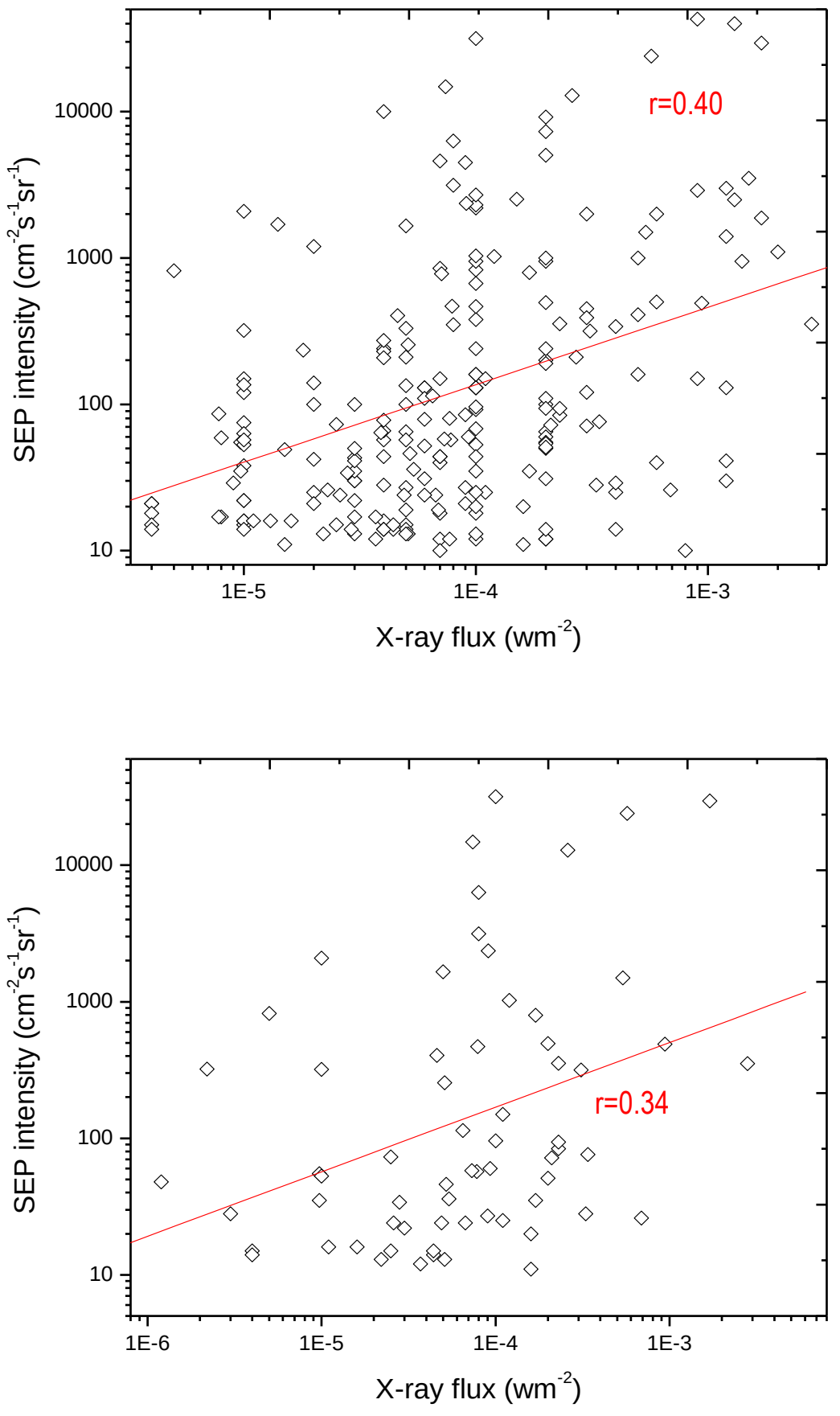

Figure 6. Scatter plots between X-ray peak flux and SEP intensity (top) and same plot for the events associated with halo CMEs (bottom). The regression line and correlation coefficients are also shown in figure.

## 4. Summary

In this paper we have studied the SEP events associated with solar X-ray flares and CMEs. For this study, a large data was taken from solar cycle 21 to 24.

We have studied the flare size, CME speed, CME width, source location with the number of SEP events and SEP intensity. The main results are summarized as follows:

— Average CME speed over that large period of data comes to be 1238 kms$^{-1}$. Maximum number of SEP events has the CME speed between 1000 kms$^{-1}$ to 1600 kms$^{-1}$.

— Average CME width is 253 degree i.e., maximum of the SEP events are associated with halo CMEs and partial halo (> 120$^{0}$) CMEs. This result confirms the previous findings that wider CMEs can produce more strong SEPs.

— Majority of SEP producing solar flares are GOES M and X-class.

— The correlation study of SEP and CME speed shows the value of correlation coefficient is 0.41 in eastern hemisphere (> 20$^{0}$ east). When we move towards the western hemisphere the coefficient becomes stronger (0.52 to 0.53). The correlation between SEP intensity and X-ray flux is found 0.40. It becomes rather poor when we take the events associated with halo CMEs only. Correlation reduces to 0.34.

— The GLE events during solar cycle 23 & 24 are associated with major flares and high speed CMEs.

— This study is interesting to demonstrate how the forecast using the characteristics of CMEs cannot be done with a better rate than 50%. Progresses on the fundamental physics should be done before having a more precise forecast.

From our analysis, we have found neither the flare intensity nor the CME speed are tightly correlated with the SEP intensity. For the better understanding of SEPs, we need to consider the magnetic topology of the active

**Table 1.** SEP events of solar cycle 23 & 24 associated with GLE.

| Date | CME speed | CME width | Source location | Flare class | SEP intensity |
|---|---|---|---|---|---|
| 06-11-97 | 1556 | H | S18W63 | X9.4 | 490 |
| 02-05-98 | 938 | H | S15W15 | X1.1 | 150 |
| 06-05-98 | 1099 | 190 | S11W65 | X2.7 | 210 |
| 24-08-98 | No CME data | | N30E07 | X1.0 | 670 |
| 14-07-00 | 1674 | H | N22W07 | X5.7 | 24000 |
| 15-04-01 | 1199 | 167 | S20W85 | X14 | 951 |
| 18-04-01 | 2465 | H | S23W92 | Filament eruption | 321 |
| 04-11-01 | 1810 | H | N06W18 | X1.0 | 31700 |
| 26-12-01 | 1446 | 212 | N08W54 | M7.1 | 779 |
| 24-08-02 | 1878 | H | S02W81 | X3.1 | 317 |
| 28-10-03 | 2459 | H | S16E08 | X17 | 29500 |
| 02-11-03 | 2598 | H | S14W56 | X8.3 | 1570 |
| 17-05-12 | 1582 | H | N11W76 | M5.1 | 255 |

regions together with the CME and flare characteristics.

Here, we would like to mention that the big uncertainty due to the measurement at L1 (a single point) limits the statistical analysis. For the better statistical results, we need an armada of spacecraft to have a better view of the solar wind. Parker solar probe and Solar Orbiter will bring certainly important information.

## Acknowledgements

We would like to thank the reviewers for their constructive and important suggestions, which improve the paper considerably. We acknowledge the open data policy of NGDC, SOHO and SDO. RC acknowledges the support from SERB-DST project no. SERB/F/7455/2017-17.

# Appendix (Data set)

| SEP | | CME | | | Source Location | Active Region No. | Flare Class | SEP Intensity(pfu) |
|---|---|---|---|---|---|---|---|---|
| Date | Time | Time | Speed | Width | | | | |
| **Data from 1976 to 1996** | | | | | | | | |
| 30-04-76 | 21:20 | **No CME Observation during 1976- 1996** | | | S09W47 | 700 | X2 | 12 |
| 19-09-77 | 14:30 | | | | N08W58 | 889 | X2 | 200 |
| 22-11-77 | 14:00 | | | | N24W38 | 939 | X1 | 160 |
| 13-02-78 | 9:30 | | | | N22W13 | 1001 | M7 | 850 |
| 11-04-78 | 15:30 | | | | N19W54 | 1057 | X2 | 65 |
| 29-04-78 | 4:45 | | | | N22E41 | 1092 | X5 | 1000 |
| 07-05-78 | 4:20 | | | | N22W64 | 1095 | X2 | 100 |
| 02-06-78 | 7:30 | | | | N23W50 | 1129 | M5 | 19 |
| 24-06-78 | 9:00 | | | | N19E18 | 1164 | M2 | 25 |
| 13-07-78 | 3:00 | | | | NO FLARE | | | 20 |
| 23-09-78 | 10:35 | | | | N35W50 | 1294 | X1 | 2200 |
| 10-11-78 | 21:30 | | | | N17E02 | 1385 | M1 | 38 |
| 17-02-79 | 20:20 | | | | N15E48 | 1574 | X2 | 31 |
| 03-04-79 | 16:00 | | | | NO FLARE | | | 45 |
| 06-06-79 | 18:50 | | | | N20E16 | 1781 | X2 | 950 |
| 07-07-79 | 0:15 | | | | NO FLARE | | | 50 |
| 19-08-79 | 8:50 | | | | N10E90 | 1943 | X6 | 500 |
| 15-09-79 | 15:00 | | | | N10E90 | 1994 | X2 | 60 |
| 16-11-79 | 4:30 | | | | N34W25 | 2110 | M1 | 75 |
| 06-02-80 | 13:40 | | | | NO FLARE | | | 12 |
| 17-07-80 | 23:00 | | | | S12E06 | 2562 | M3 | 100 |
| 30-03-81 | 9:00 | | | | N13W74 | 2993 | M3 | 30 |
| 10-04-81 | 17:45 | | | | N09W40 | 3025 | X2 | 50 |
| 24-04-81 | 15:15 | | | | N18W50 | 3049 | X5 | 160 |
| 09-05-81 | 12:00 | | | | N09E37 | 3099 | M7 | 150 |
| 15-05-81 | 3:00 | | | | N11E58 | 3106 | X1 | 130 |
| 20-07-81 | 14:30 | | | | S26W75 | 3204 | M5 | 100 |
| 25-07-81 | 6:00 | | | | NO FLARE | | | 18 |
| 10-08-81 | 1:15 | | | | S10E24 | 3257 | M4 | 57 |
| 08-10-81 | 12:35 | | | | S19E88 | 3390 | X3 | 2000 |
| 10-12-81 | 5:45 | | | | N12W16 | 3496 | M5 | 65 |
| 31-10-82 | 0:55 | | | | S13E19 | 3576 | X1 | 830 |
| 06-06-82 | 2:45 | | | | S09E72 | 3763 | X8 | 10 |
| 09-06-82 | 0:40 | | | | S11E26 | 3763 | X12 | 30 |
| 11-07-82 | 7:00 | | | | N17E73 | 3804 | X9 | 2900 |

| | | | | | | |
|---|---|---|---|---|---|---|
| 22-07-82 | 20:30 | | N29W86 | 3804 | M4 | 240 |
| 05-09-82 | 22:05 | | N11E30 | 3886 | M4 | 66 |
| 22-11-82 | 19:40 | | S11W43 | 3994 | M7 | 40 |
| 26-11-82 | 6:05 | | S11W87 | 3994 | X4 | 25 |
| 08-12-82 | 0:10 | | S14W81 | 4007 | X2 | 1000 |
| 17-12-82 | 18:45 | | S10E24 | 4026 | X12 | 130 |
| 19-12-82 | 19:20 | | N10W75 | 4022 | M9 | 85 |
| 27-12-82 | 6:00 | | S14E31 | 4033 | X2 | 190 |
| 03-02-83 | 12:00 | | S19W08 | 4077 | X4 | 340 |
| 15-06-83 | 4:35 | | S09W90 | 4201 | NO FLARE | 18 |
| 16-02-84 | 9:15 | | S12W90 | 4408 | | 660 |
| 19-02-84 | 13:10 | | N16E82 | 4421 | X2 | 55 |
| 13-03-84 | 14:40 | | NO FLARE | | | 10 |
| 14-03-84 | 4:05 | | S12W42 | 4433 | M2 | 100 |
| 25-04-84 | 13:30 | | S12E43 | 4474 | X13 | 2500 |
| 24-05-84 | 10:45 | | S09E24 | 4492 | M6 | 31 |
| 31-05-84 | 13:15 | | S08W38 | 4492 | M1 | 15 |
| 22-01-85 | 4:15 | | N01W76 | 4617 | X4 | 14 |
| 25-04-85 | 14:30 | | N02E01 | 4647 | X1 | 160 |
| 09-07-85 | 2:35 | | S16W36 | 4671 | M2 | 140 |
| 06-02-86 | 9:25 | | S04W06 | 4711 | X1 | 130 |
| 14-02-86 | 11:55 | | N01W76 | 4713 | M6 | 130 |
| 06-03-86 | 18:35 | | N02E01 | 4717 | C4 | 21 |
| 04-05-86 | 12:55 | | N06W90 | 4717 | M1 | 16 |
| 06-02-86 | 9:25 | | S04W06 | 4711 | X1 | 130 |
| 14-02-86 | 11:55 | | N01W76 | 4713 | M6 | 130 |
| 06-03-86 | 18:35 | | N02E01 | 4717 | C4 | 21 |
| 04-05-86 | 12:55 | | N06W90 | 4717 | M1 | 16 |
| 08-11-87 | 2:00 | | N31W90 | 4875 | M1 | 120 |
| 02-01-88 | 23:25 | | S34W18 | 4912 | X1 | 92 |
| 25-03-88 | 22:25 | | N22W90 | 4965 | NO FLARE | 58 |
| 30-06-88 | 10:55 | | S16E22 | 5060 | M9 | 21 |
| 26-08-88 | 0:00 | | N24E90 | 5125 | M2 | 42 |
| 12-10-88 | 9:20 | | S20W66 | 5175 | X2 | 12 |
| 08-11-88 | 22:25 | | S17W47 | 5212 | M3 | 13 |
| 14-11-88 | 1:30 | | S23W27 | 5227 | M3 | 13 |
| 17-12-88 | 6:10 | | N27E59 | 5278 | X1 | 18 |
| 17-12-88 | 20:00 | | N26E37 | 5278 | X4 | 29 |
| 04-01-89 | 23:05 | | S20W60 | 5303 | M4 | 28 |
| 08-03-89 | 17:35 | | N35E69 | 5395 | X15 | 3500 |
| 17-03-89 | 18:55 | | N33W60 | 5395 | X6 | 2000 |
| 23-03-89 | 20:40 | | N18W28 | 5409 | X1 | 53 |
| 11-04-89 | 14:35 | | N35E29 | 5441 | X3 | 450 |
| 05-05-89 | 9:05 | | S20W36 | 5464 | M5 | 27 |
| 06-05-89 | 2:35 | | N30E01 | 5470 | X2 | 110 |
| 23-05-89 | 11:35 | | N20W35 | 5477 | X1 | 68 |
| 24-05-89 | 7:30 | | S21E16 | 5497 | M5 | 15 |
| 18-06-89 | 16:50 | | N12W31 | 5534 | C4 | 18 |
| 30-06-89 | 6:55 | | N26W60 | 5555 | M3 | 17 |
| 01-07-89 | 6:55 | | N11E20 | 5590 | M3 | 17 |
| 25-07-89 | 9:00 | | N25W84 | 5603 | X2 | 54 |
| 12-08-89 | 16:00 | | S16W37 | 5629 | X2 | 9200 |
| 04-09-89 | 1:20 | | S18E16 | 5669 | X1 | 44 |
| 12-09-89 | 19:35 | | S18W79 | 5669 | M5 | 57 |
| 29-09-89 | 12:05 | | S26W90 | 5698 | X9 | 4500 |
| 06-10-89 | 0:50 | | S20E12 | 5710 | X1 | 22 |
| 19-10-89 | 13:05 | | S27E10 | 5747 | X13 | 40000 |
| 09-11-89 | 2:40 | | N18W50 | 5155 | M3 | 43 |
| 15-11-89 | 7:35 | | N11W26 | 5786 | X3 | 71 |
| 27-11-89 | 20:00 | | N30E05 | 5800 | X1 | 380 |
| 30-11-89 | 13:45 | | N26W59 | 5800 | X2 | 7300 |
| 19-03-90 | 7:05 | | N31W43 | 5969 | X1 | 950 |
| 29-03-90 | 9:15 | | S04W37 | 5988 | M4 | 16 |
| 07-04-90 | 22:40 | | N22E72 | 6007 | M7 | 18 |
| 11-04-90 | 21:20 | | N21W24 | 6013 | C3 | 13 |
| 17-04-90 | 5:00 | | N32E57 | 6022 | X1 | 12 |
| 28-04-90 | 10:05 | | S40W13 | 6050 | M1 | 150 |
| 21-05-90 | 23:55 | | N35W36 | 6063 | X5 | 410 |
| 24-05-90 | 21:25 | | N33W78 | 6063 | X9 | 150 |
| 28-05-90 | 7:15 | | NO FLARE | | | 45 |
| 12-06-90 | 11:40 | | N10W33 | 6089 | M6 | 79 |
| 26-07-90 | 17:20 | | NO FLARE | | | 21 |
| 01-08-90 | 0:05 | | N20E45 | 6180 | M4 | 230 |
| 30-01-91 | 11:30 | | S17W35 | 6469 | X1 | 240 |
| 25-02-91 | 12:10 | | S16W80 | 6497 | X1 | 13 |
| 23-03-91 | 8:20 | | S26E28 | 6555 | X9 | 43000 |
| 29-03-91 | 21:20 | | N19W50 | 6560 | X1 | 20 |
| 03-04-91 | 8:15 | | N14W00 | 6562 | M6 | 52 |
| 13-05-91 | 3:00 | | S09W90 | 6615 | M8 | 350 |
| 31-05-91 | 12:25 | | NO FLARE | | | 22 |
| 04-06-91 | 8:20 | | N30E70 | 6659 | X12 | 3000 |
| 14-06-91 | 23:40 | | N33W69 | 6659 | X12 | 1400 |

| | | | | | | | | |
|---|---|---|---|---|---|---|---|---|
| 30-06-91 | 7:55 | | | | N30E85 | 6703 | M6 | 110 |
| 07-07-91 | 4:55 | | | | N26E03 | 6703 | X1 | 2300 |
| 11-07-91 | 2:40 | | | | S22E34 | 6718 | M3 | 30 |
| 11-07-91 | 22:55 | | | | N30W44 | 6770 | X2 | 14 |
| 26-08-91 | 17:40 | | | | N25E64 | 6805 | X2 | 240 |
| 01-10-91 | 17:40 | | | | S21E32 | 6853 | M7 | 12 |
| 28-10-91 | 13:00 | | | | S13E15 | 6891 | X6 | 40 |
| 30-10-91 | 7:45 | | | | S08W25 | 6891 | X2 | 94 |
| 07-02-92 | 6:45 | | | | S13W10 | 7042 | M4 | 78 |
| 16-03-92 | 8:40 | | | | S14E29 | 7100 | M7 | 10 |
| 09-05-92 | 10:05 | | | | S26E08 | 7154 | M7 | 4600 |
| 25-06-92 | 20:45 | | | | N09W67 | 7205 | X3 | 390 |
| 06-08-92 | 11:45 | | | | S09E68 | 7248 | M4 | 14 |
| 30-10-92 | 19:20 | | | | S22W61 | 7321 | X1 | 2700 |
| 04-03-93 | 15:05 | | | | S14W56 | 7434 | C8 | 17 |
| 12-03-93 | 20:10 | | | | S00W51 | 7440 | M7 | 44 |
| 20-02-94 | 3:00 | | | | N09W02 | 7671 | M4 | 10000 |
| 20-10-94 | 0:30 | | | | N12W24 | 7790 | M3 | 35 |
| 20-10-95 | 8:20 | | | | S09W55 | 7912 | M1 | 63 |
| **Data from 1977 to 2017, where CME data is available** | | | | | | | | |
| 04-11-97 | 7:00 | 6:10 | 785 | H | S14W33 | 8100 | X2.1 | 72 |
| 06-11-97 | 13:00 | 12:10 | 1556 | H | S18W63 | 8100 | X9.4 | 490 |
| 20-04-98 | 11:00 | 10:07 | 1863 | 165 | S43W90 | - | M1.4 | 1700 |
| 02-05-98 | 14:00 | 14:06 | 938 | H | S15W15 | 8210 | X1.1 | 150 |
| 06-05-98 | 8:00 | 8:29 | 1099 | 190 | S11W65 | 8210 | X2.7 | 210 |
| 09-05-98 | 5:00 | 3:35 | 2331 | 178 | >SW90 | 8210 | M7.7 | 12 |
| 24-08-98 | 23:55 | NO DATA | | | N30E07 | 8307 | X1 | 670 |
| 25-09-98 | 0:10 | | | | N18E09 | 8340 | M7 | 44 |
| 30-09-98 | 15:20 | | | | N23W81 | 8340 | M2 | 1200 |
| 08-11-98 | 2:45 | 1:25 | 350 | 20 | NO FLARE | | | 11 |
| 14-11-98 | 8:10 | NO DATA | | | N28W90 | 8375 | C1 | 310 |
| 23-01-99 | 11:05 | | | | N27E90 | - | M5 | 14 |
| 24-04-99 | 15:00 | 13:31 | 1495 | H | >NW90 | No Flare | | 32 |
| 03-05-99 | 13:00 | 6:06 | 1584 | H | N15E32 | 8530 | M4.4 | 14 |
| 05-05-99 | 18:20 | 16:26 | 649 | 42 | N15E32 | 8525 | M4 | 14 |
| 01-06-99 | 20:00 | 19:37 | 1772 | H | >NW90 | - | C1.2 | 48 |
| 02-06-99 | 2:45 | 4:50 | 422 | 39 | NO FLARE | | | 48 |
| 04-06-99 | 8:00 | 7:26 | 2230 | 150 | N17W69 | 8552 | M3.9 | 64 |
| 18-02-00 | 10:00 | 9:54 | 890 | 118 | >NW90 | NO FLARE | | 13 |
| 04-04-00 | 17:00 | 16:32 | 1188 | H | N16W66 | 8933 | C9.7 | 55 |
| 06-06-00 | 19:00 | 15:54 | 1119 | H | N20E18 | 9026 | X2.3 | 84 |
| 10-06-00 | 18:00 | 17:08 | 1108 | H | N22W38 | 9026 | M5.2 | 46 |
| 14-07-00 | 11:00 | 10:54 | 1674 | H | N22W07 | 9077 | X5.7 | 24000 |
| 22-07-00 | 12:00 | 11:54 | 1230 | 105 | N14W56 | 9085 | M3.7 | 17 |
| 28-07-00 | 10:50 | 16:54 | 394 | 9 | NO FLARE | | | 18 |
| 11-08-00 | 16:50 | 16:54 | 300 | 35 | | | | 17 |
| 12-09-00 | 13:00 | 11:54 | 1550 | H | S17W09 | 9163 | M1 | 320 |
| 16-10-00 | 8:00 | 7:27 | 1336 | H | >W90 | 9193 | M2.5 | 15 |
| 25-10-00 | 12:00 | 8:26 | 770 | H | N10W66 | 9199 | C4 | 15 |
| 08-11-00 | 23:00 | 23:06 | 1345 | H | N10W77 | 9213 | M7.4 | 14800 |
| 24-11-00 | 14:00 | 15:30 | 1245 | H | N22W07 | 9230 | X2.3 | 94 |
| 28-01-01 | 17:00 | 15:54 | 916 | 250 | S04W59 | 9313 | M1.5 | 49 |
| 29-03-01 | 11:00 | 10:26 | 942 | H | N20W19 | 9393 | X1.7 | 35 |
| 02-04-01 | 23:00 | 22:06 | 2505 | 244 | N19W72 | 9390 | X20 | 1100 |
| 10-04-01 | 8:00 | 5:30 | 2411 | H | S23W09 | 9415 | X2.3 | 355 |
| 12-04-01 | 12:00 | 10:31 | 1184 | H | S19W43 | 9415 | X2 | 51 |
| 15-04-01 | 14:00 | 14:06 | 1199 | 167 | S20W85 | 9415 | X14 | 951 |
| 18-04-01 | 3:00 | 2:30 | 2465 | H | S23W92 | 9424 | C2.2 | 321 |
| 28-04-01 | 14:00 | 12:30 | 1006 | H | N17W31 | 9433 | M7.8 | 57 |
| 07-05-01 | 13:00 | 12:06 | 1223 | 205 | >NW90 | NO FLARE | | 30 |
| 15-06-01 | 16:00 | 15:56 | 1701 | H | >SW90 | | | 26 |
| 09-08-01 | 19:00 | NO DATA | | | S17E19 | 9570 | C7.8 | 17 |
| 16-08-01 | 1:00 | 23:54 | 1575 | H | Backside | NO FLARE | | 493 |
| 15-09-01 | 2:00 | 11:54 | 478 | 130 | S21W49 | 9608 | M1.5 | 11 |
| 24-09-01 | 11:00 | 10:30 | 2402 | H | S16E23 | 9632 | X2.6 | 12900 |
| 01-10-01 | 13:00 | 5:30 | 1405 | H | S20W84 | 9628 | M9.1 | 2360 |
| 19-10-01 | 17:30 | 16:50 | 901 | H | N15W29 | 9661 | X1.6 | 11 |
| 22-10-01 | 17:00 | 15:06 | 1336 | H | S21E18 | 9672 | M6.7 | 24 |
| 04-11-01 | 17:00 | 16:35 | 1810 | H | N06W18 | 9684 | X1 | 31700 |
| 17-11-01 | 6:00 | 5:30 | 1380 | H | S13E42 | 9704 | M2.8 | 34 |
| 22-11-01 | 21:00 | 20:30 | 1443 | H | S25W67 | 9704 | M3.8 | |
| 22-11-01 | 24:00:00 | 23:30 | 1437 | H | S15W34 | 9704 | M9.9 | |
| 26-12-01 | 5:30 | 5:30 | 1406 | 212 | N08W54 | 9742 | M7.1 | 779 |
| 29-12-01 | 24:00:00 | 20:06 | 2044 | H | S26E90 | 9756 | X3.4 | 76 |
| 08-01-02 | 3:00 | 17:54 | 1794 | H | >NE90 | NO FLARE | | 91 |
| 14-01-02 | 24:00:00 | 5:35 | 1492 | H | S28W83 | - | M4.4 | 15 |
| 20-02-02 | 6:00 | 6:30 | 952 | H | N12W72 | 9825 | M5.1 | 13 |
| 15-03-02 | 3:00 | 23:06 | 907 | H | S08W03 | 9866 | M2.2 | 13 |
| 18-03-02 | 6:00 | 2:54 | 989 | H | S09W46 | 9866 | M1 | 53 |
| 22-03-02 | 13:30 | 11:06 | 1750 | H | >SW90 | 9986 | M1.6 | 16 |
| 17-04-02 | 10:30 | 8:26 | 1218 | H | S14W34 | 9906 | M2.6 | 24 |
| 21-04-02 | 2:30 | 1:27 | 2409 | 241 | S14W84 | 9906 | X1.5 | 2520 |
| 22-05-02 | 6:00 | 3:50 | 1494 | H | S30W34 | - | C5 | 820 |
| 07-07-02 | 13:00 | 11:06 | 1329 | >205 | >W90 | 17 | M1 | 22 |
| 15-07-02 | 10:30 | 21:30 | 1300 | >188 | N19W01 | 30 | M1.8 | 234 |
| 20-07-02 | 6:00 | 21:30 | 2017 | H | >SE90 | 39 | X3.3 | 28 |
| 14-08-02 | 3:00 | 2:30 | 1309 | 133 | N09W54 | 61 | M2.3 | 26 |
| 22-08-02 | 2:30 | 2:06 | 1005 | H | S07W62 | 69 | M5.4 | 36 |
| 24-08-02 | 1:30 | 1:27 | 1878 | H | S02W81 | 69 | X3.1 | 317 |
| 07-09-02 | 4:40 | 3:54 | 668 | 66 | N09E28 | 102 | M4 | 208 |
| 09-11-02 | 15:00 | 13:31 | 1838 | H | S12W29 | 180 | M4.6 | 404 |
| 28-05-03 | 23:35 | 00:50 | 1366 | H | S07W17 | 365 | X3 | 121 |
| 31-05-03 | 4:40 | 2:30 | 1835 | H | S07W65 | 365 | M9 | 27 |
| 18-06-03 | 20:50 | 13:31 | 586 | 135 | S08E61 | 386 | M6 | 24 |
| 26-10-03 | 18:25 | 17:54 | 1537 | 171 | N02W38 | 484 | X1 | 466 |
| 28-10-03 | 12:15 | 11:30 | 2459 | H | S16E08 | 486 | X17 | 29500 |
| 02-11-03 | 11:05 | 11:30 | 826 | 33 | NO FLARE | | | 1570 |
| 04-11-03 | 22:25 | 19:54 | 2657 | H | S19W83 | 486 | X28 | 353 |
| 21-11-03 | 23:55 | 19:27 | 737 | 82 | N02W17 | 501 | M5 | 13 |
| 02-12-03 | 15:05 | 10:50 | 1393 | 150 | W limb | - | C7.8 | 86 |
| 11-04-04 | 11:35 | 11:54 | 1132 | H | S14W47 | 588 | C9.7 | 35 |
| 25-07-04 | 18:55 | 14:54 | 1333 | H | N08W33 | 652 | M1 | 2086 |
| 13-09-04 | 21:05 | 20:48 | 289 | 47 | N04E42 | 672 | M4 | 273 |
| 19-09-04 | 19:25 | 22:18 | - | 99 | N03W58 | 672 | M1 | 57 |
| 01-11-04 | 6:55 | 6:06 | 925 | 146 | Backside | NO FLARE | | 63 |
| 07-11-04 | 19:10 | 16:54 | 1759 | H | N09W17 | 696 | X2 | 495 |
| 16-01-05 | 2:10 | 5:30 | 765 | 31 | N15W05 | 720 | X2 | 5040 |
| 14-05-05 | 5:25 | 17:12 | 1689 | H | N12E11 | 759 | M8 | 3140 |
| 16-06-05 | 22:00 | 13:36 | 424 | 85 | N09W87 | 775 | M4 | 44 |
| 14-07-05 | 2:45 | 3:54 | 259 | 25 | N10W80 | 786 | M5 | 134 |
| 27-07-05 | 23:00 | 23:54 | 246 | 21 | N11E90 | 792 | M3 | 41 |
| 22-08-05 | 20:40 | 22:30 | 819 | 57 | S12W60 | 798 | M5 | 330 |
| 08-09-05 | 2:15 | No Data | | | S06E89 | 808 | X17 | 1880 |
| 14-08-10 | 12:30 | 10:12 | 1250 | H | N17W52 | 1099 | C4 | 14 |
| 07-03-11 | 21:50 | 20:00 | 2125 | H | N30W47 | 1164 | M3.7 | 12 |
| 08-03-11 | 1:05 | 20:15 | 712 | 99 | N24W59 | 1164 | M3 | 50 |
| 21-03-11 | 19:50 | 2:24 | 1341 | H | Backside | 1169 | NO FLARE | 14 |
| 07-06-11 | 8:20 | 6:49 | 1255 | H | S21W54 | 1226 | M2.5 | 73 |
| 04-08-11 | 5:16 | 4:12 | 1315 | H | N19W36 | 1261 | M9.3 | 60 |
| 09-08-11 | 8:45 | 8:12 | 1610 | H | N17W69 | 1263 | X6.9 | 26 |
| 23-09-11 | 22:55 | 0:48 | 1116 | 44 | N11E74 | 1302 | X1/2N | 35 |
| 26-11-11 | 11:25 | 7:12 | 933 | H | N08W49 | 1353 | NO FLARE | 80 |
| 23-01-12 | 5:30 | 4:00 | 2175 | H | N28W36 | 1402 | M8 | 6310 |
| 27-01-12 | 18:55 | 18:27 | 2508 | H | N27W71 | 1402 | X1.7 | 795 |
| 07-03-12 | 2:50 | 0:24 | 2684 | H | N17E27 | 1429 | X5.4 | 1500 |
| 13-03-12 | 18:05 | 17:36 | 1554 | H | N17W66 | 1429 | M7.9 | 469 |
| 17-05-12 | 1:55 | 1:48 | 1582 | H | N11W76 | 1476 | M5.1 | 255 |
| 27-05-12 | 5:35 | 5:48 | 725 | 197 | NO FLARE | | | 14 |
| 16-06-12 | 19:55 | 15:48 | 151.00 | 10 | S17E14 | 1504 | M1 | 14 |
| 06-07-12 | 5:00 | 23:24 | 1828 | H | S13W59 | 1515 | X1.1 | 25 |
| 07-07-12 | 4:00 | 4:48 | 9 | 195 | S18W50 | 1515 | X1 | 25 |
| 08-07-12 | 18:10 | 16:54 | 1495 | 157 | S17W74 | 1515 | M6.9 | 19 |
| 12-07-12 | 18:35 | 16:48 | 885 | H | S16W09 | 1520 | X1 | 96 |
| 17-07-12 | 17:15 | 17:00 | 395 | 96 | S17W75 | 1520 | M1 | 136 |
| 19-07-12 | 6:40 | 4:24 | 444 | 11 | S13W88 | 1520 | M7.7 | 80 |
| 23-07-12 | 15:45 | 17:48 | 411 | 27 | S16W86 | 1520 | NO FLARE | 12 |
| 01-09-12 | 13:35 | 13:36 | 511 | 35 | S06E20 | - | C8 | 59 |
| 28-09-12 | 3:00 | 0:12 | 947 | H | N08W41 | 1577 | C3 | 28 |
| 15-03-13 | 19:40 | 7:12 | 1063 | H | N11E12 | 1692 | M1.1 | 16 |
| 11-04-13 | 8:25 | 7:24 | 861 | H | N09E12 | 1719 | M6.5 | 114 |
| 14-05-13 | 13:25 | 12:36 | 425 | 10 | N11E51 | 1748 | X12 | 41 |
| 22-05-13 | 14:20 | 13:25 | 1466 | H | N15W70 | 1745 | M5 | 1660 |
| 23-06-13 | 8:30 | 21:24 | 339 | 101 | S15E62 | 1778 | M2.9 | 14 |
| 30-09-13 | 5:05 | 13:25 | 1466 | H | N15W40 | N/A | N/A | 182 |
| 28-12-13 | 21:50 | 22:12 | 372 | 42 | S18E07 | 1936 | C9 | 29 |
| 06-01-14 | 9:15 | 10:24 | 957 | 76 | S15E90 | 1936 | N/A | 42 |
| 06-01-14 | 9:15 | 10:24 | 957 | 76 | S15W11 | 1944 | X1 | 1033 |
| 07-01-14 | 19:55 | 18:24 | 1830 | H | S15W11 | 1944 | X1.2 | 1026 |
| 20-02-14 | 8:15 | 8:04 | 948 | H | S15W73 | 1976 | M3 | 22 |
| 25-02-14 | 3:50 | 1:25 | 2147 | H | S12E62 | 1990 | M4.9 | 24 |
| 18-04-14 | 13:40 | 13:31 | 1203 | H | S20W34 | 2036 | NM7.3 | 58 |
| 10-09-14 | 21:35 | 18:00 | 1425 | H | N14E02 | 2158 | X1.6 | 20 |
| 18-06-15 | 11:35 | 9:36 | 189 | 35 | S13W102 | 2365 | M1.3 | 16 |
| 29-10-15 | 5:50 | 2:36 | 530 | 202 | NO FLARE | | | 23 |
| 02-01-16 | 4:30 | 4:12 | 271 | 29 | S21W73 | 2473 | M2 | 21 |
| 05-09-17 | 7:51 | 15:42 | 377 | 47 | S08W16 | 2673 | M5 | 210 |